\documentclass[epjST]{svjour}
\usepackage{graphics}

\newcommand{\be}{\begin{equation}}
\newcommand{\ee}{\end{equation}}
\newcommand{\bea}{\begin{eqnarray}}
\newcommand{\eea}{\end{eqnarray}}

\usepackage{amssymb,cite,graphicx}
\usepackage{slashed}
\usepackage{amsmath,bm,bbm}
\usepackage{amsfonts}
\begin{document}
\title{Dark mesons as self-interacting dark matter}
\author{Hyun Min Lee\,\thanks{\email{hminlee@cau.ac.kr}}  }
\institute{Department of Physics, Chung-Ang University, Seoul
      06974, Korea}
\abstract{
We review the current status of model building for light dark matter in theories of QCD-like gauge groups in the hidden sector. The focus is upon the dark mesons with the $SU(3)_V$ flavor symmetry in scenarios of Strongly Interacting Massive Particles. We show  the production mechanism and the kinetic equilibrium condition for dark mesons and discuss a unitarization of dark chiral perturbation theory with vector mesons in the scheme of hidden gauge symmetry.
} 
\maketitle
\section{Introduction}

Dark flavor or gauge symmetries have been playing important roles for model building for dark matter beyond the minimal scenarios with a single component dark matter.
In particular, the accidental flavor symmetry in the dark sector ensures the stability of dark matter with multiple components, naturally rendering dark matter self-interacting to solve the small-scale problems at galaxies \cite{smallscale,smallscale2}. 
We focus on the contact self-interactions for dark mesons in this review, but we note that there are alternative ways to make dark matter self-interacting through the non-perturbative enhancement with light mediators \cite{lightmed}. We also remark that the effects of baryons and supernova feedback in simulations could resolve the small-scale problems \cite{baryons}, although there are issues such as diversity problem that disk galaxies with the same maximal circular velocity exhibit a much larger scatter \cite{diversity}. 

In the case of dark flavor symmetry, dark quarks form meson bound states after the condensation of hidden QCD, which are natural candidates for light dark matter due to small masses for dark quarks. The dark flavor symmetry determines the self-interactions for dark mesons in the dark chiral perturbation theory (ChPT) and allows for the Wess-Zumino-Witten (WZW) term with a coefficient fixed by the number of colors  for a sufficiently large number of dark mesons. Then, we can determine the relic density for dark mesons by the freeze-out process with $3\to 2$ meson-number changing processes \cite{3to2,simpmeson1,simpmeson2,simpmeson3}. This is the so called  Strongly Interacting Massive Particles (SIMPs), which should be distinguished from the case of dark matter candidates interacting with QCD. Then, the dark mesons have masses of sub-GeV scale for the correct relic density and they have a naturally large self-cross scattering cross section due to self-interactions to solve the small-scale problems. 

The dark flavor symmetry can be broken partially by the gauging with local dark symmetries. When the dark flavor symmetry is gauged by a local $U(1)'$, the gauge kinetic mixing between the $U(1)'$ and the hypercharge in the Standard Model (SM) communicates between the dark matter and the SM, so it is important to look for light resonances in the searches for SIMP dark matter at the intensity frontier experiments.
Furthermore, when the dark flavor symmetry is gauged by a broken local non-abelian gauge symmetry, the so called hidden gauge symmetry, we can realize vector mesons in the dark sector as excitations in the dark ChPT and extend the dark ChPT for a better behavior at high energies. 

Although we are not pursuing in this short review, we also remark the case of dark gauge symmetry. When dark gauge symmetries are broken spontaneously by Higgs mechanism, they can make the massive gauge bosons stable due to the remaining custodial symmetry from the scalar potential. 
Non-abelian gauge symmetries such as $SU(2)$ in the dark sector \cite{vsimp,vsimp2} fixe the self-interactions for dark gauge bosons in terms of the dark gauge coupling only. As a result, there is a definite prediction for the masses and gauge coupling for dark gauge bosons, being compatible with the relic density and the Bullet cluster bound.
It is worthwhile to mention that the Higgs phase for the (partially) broken dark gauge symmetry with no light charged particles is dual to the confining phase for the hidden QCD under the name of Higgs/QCD complementarity \cite{vsimp}. 
As a result, we could have the unified picture of strongly interacting dark matter as branching out from various realizations of hidden QCD with or without light dark quarks.

\section{Dark QCD and dark matter}

Suppose that the flavor symmetry for dark fermions, $G=SU(N_f)_L\times SU(N_f)_R$ is broken to the diagonal subgroup, $H=SU(N_f)_V$, due to the $SU(N_c)$ condensate of dark fermions. Then, dark mesons made of dark fermions, appear as pseudo-Goldstone bosons, and they are naturally strongly interacting, light and stable due to the flavor symmetry.
The number of dark mesons is given by the number of broken generators of the flavor symmetry, which is $N^2_f-1$. In the case of QCD, $N_c=3$ and $N_f=3$, but there are more possibilities for dark sector with arbitrary numbers of colors and flavors. 

The flavor symmetry and the  dark QCD group can be generalized to other Lie groups such as $G/H=SU(N_f)/SO(N_f)$ for $SO(N_c)$ and $SU(2N_f)/Sp(2N_f)$ for $Sp(N_c)$. In particular, a nonzero Wess-Zumino-Witten (WZW) term \cite{wz,witten} for dark mesons exists only for a nontrivial homotopy group, $\pi_5(G/H)=Z$, i.e. $N_f\geq 3$. 

For simplicity and illustration, we focus on the flavor symmetry, $SU(3)_L\times SU(3)_R/SU(3)_V$, in the following discussion.  Then, dark mesons are represented as $\Sigma={\rm exp}(i2\pi/f_\pi)$ with $\pi=\pi^a t^a$, in the basis of Gell-Mann matrices, with
\bea
\pi =\frac{1}{\sqrt{2}} \left(\begin{array}{ccc} \frac{1}{\sqrt{2}} \pi^0+\frac{1}{\sqrt{6}}\eta^0   & \pi^+ & K^+ \\  
\pi^-& -\frac{1}{\sqrt{2}} \pi^0+\frac{1}{\sqrt{6}}\eta^0   & K^0 \\  K^- & \overline{K^0} & -\frac{2}{\sqrt{6}}\eta^0 \end{array} \right).
\eea
Here, we note that $\eta'$ meson can be also included when the global symmetry is extended to $U(3)_L\times U(3)_R$, which is broken down to $U(3)_V$. 

When the mass matrix for dark quarks is of diagonal form,
\bea
M_q=\left(\begin{array}{ccc} m_1& 0 & 0 \\ 0 & m_2 & 0 \\ 0 & 0 & m_3 \end{array} \right), \label{darkquarks}
\eea
the corresponding masses for dark mesons are
\bea
m^2_{{\pi}^\pm} &=& \mu (m_1+m_2), \\
m^2_{{ K}^\pm} &=& \mu (m_1+m_3), \\
m^2_{{ K}^0} &=& \mu (m_2 + m_3),
\eea
and ${\pi}^0,{\eta}^0$ mix by the following mixing mass matrix,
\bea
M^2_0 = \mu \left(\begin{array}{cc} m_1+m_2 & \frac{1}{\sqrt{3}}(m_1-m_2) \\   \frac{1}{\sqrt{3}}(m_1-m_2) &  \frac{1}{3}(m_1 + m_2 + 4m_3) \end{array} \right).
\eea
For $m_1=m_2=m_3$, all the dark mesons have common masses, $m^2_\pi=2\mu m_1$, being consistent with the unbroken $SU(3)_V$ flavor symmetry. Then, the remaining flavor symmetry ensures the stability of dark mesons.

\section{The relic abundances of SIMP mesons}

The WZW term for dark mesons contain the 5-point self-interactions, given \cite{wz,witten,simpmeson1} by
\bea
{\cal L}_{WZW}=\frac{2 N_c}{15\pi^2 f^5_\pi} \,\epsilon^{\mu\nu\rho\sigma} {\rm Tr}[ \pi \partial_\mu\pi \partial_\nu \pi \partial_\rho\pi \partial_\sigma \pi]. \label{WZW}
\eea
Then, the WZW term gives rise to $3\rightarrow 2$ annihilation channels for dark mesons, so it determines the relic density for dark mesons in the early Universe after freeze-out. The corresponding annihilation cross section for $3\rightarrow 2$ processes \cite{simpmeson1,average} is given by
\bea
\langle\sigma v^2\rangle_{3\rightarrow 2} = \frac{5\sqrt{5} N^2_c m^5_\pi}{2\pi^5 f^{10}_\pi} \frac{t^2}{N^3_\pi} \, x^{-2}\equiv \frac{\alpha^3_{\rm eff}}{m^5_\pi}\, x^{-2}
\eea
where $x\equiv m_\pi/T$ with $T$ being the radiation temperature of the Universe,  $N_\pi$ is the number of dark mesons and $t^2$ is the group theory factor, given by $N_\pi=N^2_f-1$ and $t^2=\frac{4}{3} N_f (N^2_f-1)(N^2_f-4)$ for the $SU(N_f)$ flavor symmetry \cite{simpmeson1}, respectively. In our case, we take $N_\pi=8$ and $N_f=3$, for which $t^2=160$.
Henceforth, we assume that the maximum temperature of the Universe is large enough for dark mesons to be initially populated in equilibrium with the SM.

When dark mesons annihilate dominantly by the WZW term, the relic densities for dark mesons, $n_{\rm \pi_i}=\frac{1}{8} n_{\rm DM}$, are the same and they are governed by the following Boltzmann equation,
\bea
{\dot n}_{\rm DM} +3 H n_{\rm DM} =-\langle\sigma v^2\rangle_{3\rightarrow 2}  \, n^2_{\rm DM}  \Big( n_{\rm DM}-n^{\rm eq}_{\rm DM}\Big)
\eea
where $n^{\rm eq}_{\rm DM}$ is the equilibrium number density of dark matter. As a result, the dark matter relic density is determined \cite{average} to be
\bea
\Omega_{\rm DM} h^2&=& \frac{1.05\times 10^{-10}\,{\rm GeV}^{-2}}{g^{3/4}_* (6M_P m^3_\pi/\alpha^3_{\rm eff})^{-1/2}x^{-3}_f} \nonumber \\
&=&0.12 \bigg(\frac{10.75}{g_*} \bigg)^{3/4}\bigg(\frac{x_f}{15}\bigg)^3 \Big(\frac{m_\pi}{300\,{\rm MeV}} \Big)^{3/2} \Big(\frac{46}{\alpha_{\rm eff}} \Big)^{3/2}. \label{relic}
\eea
Here, the effective self-coupling is given by
\bea
\alpha_{\rm eff} =\frac{1}{N_\pi}\bigg(\frac{5\sqrt{5}N^2_c t^2}{2\pi^5} \bigg)^{1/3} \bigg(\frac{m_\pi}{f_\pi} \bigg)^{10/3}=46 \Big(\frac{N_c}{3}\Big)^{2/3} \bigg(\frac{m_\pi/f_\pi}{4.2} \bigg)^{10/3}
\eea
where we took $N_\pi=8$ and $N_f=3$ in the second equality.
For a large $N_f$, we have $N^2_c t^2/N^3_\pi\simeq  4N^2_c/(3 N_f)$, so we need a large number of colors to maintain the effective self-coupling. In order to get the correct relic density from eq.~(\ref{relic}), we need to choose $m_\pi/f_\pi$ to be close to unitarity bound, as follows,
\bea
\frac{m_\pi}{f_\pi}\simeq 4.2\bigg(\frac{m_\pi}{300\,{\rm MeV}} \bigg)^{3/10}\Big(\frac{3}{N_c}\Big)^{1/5}. \label{relicc}
\eea

On the other hand, in the dark ChPT for dark mesons, the self-scattering cross section for dark mesons is given \cite{simpmeson1}, as follows,
\bea
\frac{\sigma_{\rm self}}{m_\pi} =\frac{m_\pi a^2}{32\pi N^2_\pi f^4_\pi}
\eea
with $a^2=8(N^2_f-1)(3N^4_f-2N^2_f+6)/N^2_f$ for the $SU(N_f)$ flavor symmetry.
For $N_f=3$, we get $a^2/N^2_\pi=\frac{77}{3}$ and it becomes saturated to $24$ for a large $N_f$. 
Then, due to the bound from Bullet cluster \cite{smallscale2}, $\sigma_{\rm self}/m_\pi\lesssim 1\,{\rm cm^2/g}$, we can set the bound on the dark matter self-coupling for $N_f=3$, as follows,
\bea
\alpha_{\rm eff} \lesssim 64  \bigg(\frac{m_\pi}{300\,{\rm MeV}} \bigg)^{5/2} \Big(\frac{N_c}{3} \Big)^{2/3}. \label{selfc}
\eea
Therefore, from eqs.~(\ref{relic}) and (\ref{selfc}), we can satisfy the relic density condition and the Bullet cluster bound for $m_\pi\sim 300\,{\rm MeV}$ and $\alpha_{\rm eff}\sim 46$ (or $m_\pi/f_\pi\sim 4.2$).

\section{$Z'$ portal and kinetic equilibrium}

For communication between dark matter and the SM, we consider the partial gauging of the flavor symmetry with a dark local $U(1)'$. The corresponding gauge boson $Z'$ with mass $m_{Z'}$ has a gauge kinetic mixing with the SM hypercharge as ${\cal L}_{\rm g.m}=-\frac{1}{2} \sin\xi\, F'_{\mu\nu} B^{\mu\nu}$ with $F'_{\mu\nu}=\partial_\mu Z'_\nu-\partial_\nu Z'_\mu$ \cite{simpmeson2}. 
We take the charge operator $Q'$ for dark quarks under the $U(1)'$ \cite{simpmeson2} to be diagonal but non-universal, as follows,
\bea
Q'=\left(\begin{array}{ccc} 1& 0 & 0 \\ 0 & -1 & 0 \\ 0 & 0 & -1 \end{array} \right). \label{charge}
\eea 
Then, the AVV anomalies for the dark chiral symmetry are absent because ${\rm Tr}(Q^{\prime 2} t^a)=0$ for $t^a\in SU(3)$  for $Q^{\prime 2} =0$. Then, neutral SIMP mesons are stable. But, there exist AAAV anomaly terms for $\pi-\pi-\pi-Z'$ interactions \cite{simpmeson2}. 
In comparison, for QCD, we have $Q={\rm diag}(2/3,-1/3,-1/3)$ for electromagnetism, for which neutral mesons in QCD become unstable due to chiral anomalies.

When the mass matrix for dark quarks in eq.~(\ref{darkquarks}) is proportional to identity, the dark charge operator in eq.~(\ref{charge}) leads to the following $Z'$ gauge interactions for dark mesons \cite{simpmeson2},
\bea
{\cal L}_{Z',2\pi} &=& 2i g_{Z'} Z'_\mu \Big(K^+\partial^\mu K^- -K^- \partial^\mu K^+ + \pi^+ \partial^\mu \pi^- -  \pi^- \partial^\mu \pi^+ \Big) \nonumber \\
&&+ 4 g^2_{Z'}Z'_\mu Z^{\prime \mu} (K^+ K^- + \pi^+ \pi^-) \label{zpdm}
\eea
where $g_{Z'}$ is the $Z'$ gauge coupling. 
Moreover, for a small gauge kinetic mixing, $\xi\ll 1$, and $m_{Z'}\ll m_Z$, the $Z'$ interactions to the SM are given \cite{z3dm,exodm} by
\bea
{\cal L}_{Z',{\rm SM}}=- e\varepsilon Z'_\mu \bigg( J^\mu_{\rm EM}+\frac{m^2_{Z'}}{2c^2_W m^2_Z}\, J^\mu_Z  \bigg)
\eea
where $\varepsilon\equiv c_W \xi$ with $c_W=\cos\theta_W$, and  $J^\mu_{\rm EM}, J^\mu_Z$ are electromagnetic and neutral currents in the SM, for instance, $J^\mu_{\rm EM}={\bar e} \gamma^\mu e$ for electron and $J^\mu_{Z}={\bar \nu} \gamma^\mu P_L \nu$ for neutrinos.

The $Z'$ portal interaction is crucial to maintain dark matter in kinetic equilibrium with the SM plasma during freeze-out \cite{z3dm,simpmeson2,vsimp,exodm,vsimp2}. Otherwise, the dark matter temperature could differ from the radiation temperature, getting unsuppressed due to continuous $3\to 2$ annihilations until late times and preventing the dark matter from making the structure formation \cite{3to2}. 

The time evolution of the kinetic energy $K$ for dark mesons with $3\to 2$ annihilation processes \cite{vsimp} is dictated by
\bea
{\dot K} +2H K =-m^2_\pi H T^{-1} +T\gamma_\pi(T)
\eea
where $\gamma_\pi(T)$ is the momentum relaxation rate for dark mesons.
Then, in the case with $3\to 2$ dominance for dark matter annihilation, the kinetic equilibrium is achieved for $\gamma_\pi(T)>H (m_\pi/T)^2$.

In the presence of $Z'$ portal couplings in eq,~(\ref{zpdm}), the charged dark mesons scatter off the SM particles in the thermal plasma by  ${K}^\pm e\rightarrow { K}^\pm e$ and  ${\pi}^\pm e\rightarrow { \pi}^\pm e$.
Then, the corresponding momentum relaxation rate for the dark mesons, ${ K}^\pm$ and $\pi^\pm$ \cite{simpmeson2}, are given by
\bea
\gamma_{{ K}^\pm}=\gamma_{{ \pi}^\pm}= \frac{320 \zeta(7)}{\pi^3} \frac{\varepsilon^2 e^2 g^2_{Z'}}{ m_\pi m^4_{Z'}}\, T^6. \label{kineq}
\eea
Therefore, as far as $\gamma_{{ K}^\pm}=\gamma_{{ \pi}^\pm}>H (m_\pi/T)^2$, the dark mesons,  ${ K}^\pm$ and $\pi^\pm$, remain in kinetic equilibrium with the SM, and so do the rest neutral dark mesons due to their strong self-interactions with the charged mesons.
From $H=0.33 g^{1/2}_* T^2/M_P$ at $T=15/m_\pi$, the condition for kinetic equilibrium at freeze-out with eq.~(\ref{kineq}) becomes
\bea
|\varepsilon| g_{Z'} \gtrsim 1.4\times 10^{-4}  \Big(\frac{m_{Z'}}{1\,{\rm GeV}} \Big)^2\Big(\frac{x_f}{15}\Big)^3\bigg(\frac{300\,{\rm MeV}}{m_\pi} \bigg)^{3/2}.
\eea

We remark several issues with the introduction of the flavor-dependent $U(1)'$. 
First, the charge operator in eq.~(\ref{charge}) breaks the $SU(3)_V$ flavor symmetry to an $SU(2)_V$ subgroup, so higher dimensional operators violating the flavor symmetry for dark mesons, such as $\frac{f_\pi}{\Lambda^2}\partial_\mu \pi^a ({\bar l}\gamma^\mu l)$ or $\frac{f^2_\pi}{\Lambda^3}\, \pi^a F_{\mu\nu} {\tilde F}^{\mu\nu}$  must be sufficiently suppressed for the stability of neutral dark mesons \cite{simpmeson2}. 

Moreover, dark mesons receive mass corrections due to $Z'$ gauge interactions, so masses for  charged and uncharged dark mesons get split as $\Delta m^2_\pi=c \,g^2_{Z'}f^2_\pi $ where $c\sim \frac{1}{16\pi^2} \frac{\mu^2}{m^2_{Z'}}$. Nonetheless, the charged dark mesons can be stable dark matter as far as they are the lightest particles charged under the $U(1)'$, and their relic densities can be still determined dominantly by the $3\to 2$ processes as described for the exact dark flavor symmetry, as far as the mass splitting is small enough for the $2\to 2$ annihilation processes with dark mesons only to remain decoupled at low temperatures comparable to the mass splitting.  

Furthermore, due to the gauging of the WZW term with $U(1)'$, there are extra couplings between dark mesons and $Z'$, leading to additional annihilation channels, $\pi \pi\rightarrow \pi Z'$ and $\pi\pi\to Z'Z'$ \cite{simpmeson2}. 
Then, we need to choose $m_{Z'}>m_\pi$ in order to forbid such $2\to 2$ channels during the freeze-out. 
But, we also note that if the $Z'$ mass is close to dark meson masses, the forbidden $2\to 2$ channels can be important for determining the relic density \cite{semi}.

We refer to Ref.~\cite{zportal} for the detailed meson phenomenology with $Z'$ portal and to Refs.~\cite{semi,split,split2} for the recent developments of cosmology and phenomenology on split dark mesons.
There are alternative ways to maintain the kinetic equilibrium of dark mesons by axion-like couplings \cite{axion}.

\section{Perturbativity and vector resonances}

As we discussed in Section 3, the correct relic density is achieved when the dark chiral perturbation theory with dark mesons is close to the unitarity bound, namely, we need a large value of $m_\pi/f_\pi$.  In this section, we include the dark vector mesons in the scheme of hidden local symmetry and discuss their impacts on unitarizing the dark matter self-coupling, being compatible with the Bullet cluster bound. 

Including hidden local symmetry $H_{\rm local}=SU(3)_V$ in addition to the global symmetry $G=SU(3)_L\times SU(3)_R$,  vector mesons can be expressed in the following matrix form \cite{simpmeson3},
\begin{equation}
V_\mu(x) = \frac{1}{\sqrt{2}}~
\begin{pmatrix} 
\frac{1}{\sqrt{2}} \rho_\mu^0 + \frac{1}{\sqrt{6}} \omega_{8 \mu}
& \rho_\mu^+ & K_\mu^{*+}  
\\
\rho_\mu^- &  -\frac{1}{\sqrt{2}} \rho_\mu^0 + \frac{1}{\sqrt{6}} \omega_{8 \mu} 
 &  K_\mu^{*0} 
\\        
K_\mu^{*-} & \overline{K_\mu^{*0}} & - \frac{2}{\sqrt{6}} \omega_{8 \mu}.
\end{pmatrix} 
\end{equation}
Here, we note that $\omega_0$ vector meson can be also included when the global symmetry is extended to $U(3)_L\times U(3)_R$, which is broken down to $U(3)_V$. 
The masses and couplings of vector mesons to dark mesons are given by
\bea
\Delta {\cal L}_V  =  m_V^2 {\rm Tr} V_\mu V^\mu - 2 i g_{V\pi\pi} {\rm Tr} \left( V_\mu [ \partial^\mu \pi , \pi ]  \right) -\frac{a}{4f^2_\pi} {\rm Tr} \left([\pi,\partial_\mu \pi]^2
\right)
\eea
with
\bea
m_V^2  &=& a g^2 f_\pi^2, \label{EqMV} \\
g_{V\pi\pi}  &=&  \frac{1}{2} a g.  \label{EqG}
\eea
Due to the full flavor symmetry, $H=SU(3)_V$, vector mesons have common masses and universal interactions to dark mesons. 
In the ordinary hadron system $a \simeq 2$, but $a$ can be considered as a free parameter 
in the dark ChPT.  Then, we can take $m_V$ and $m_\pi$ to be independent
by suitably varying the dark quark masses. 

For vector mesons, the WZW term can be generalized due to gauge invariance under the hidden local symmetry, as follows,
\begin{equation}
\Gamma^{anom} = \int d^4x \left[ \mathcal{L}_{WZW} 
-15C(c_1 {\cal L}_1  + c_2 {\cal L}_2+c_3{\cal L}_3)  \right] 
\end{equation}
where  $C \equiv - i \frac{N_c}{240 \pi^2} $, and
the gauged WZW terms, ${\cal L}_{1,2,3}$, contain the interactions between vector mesons and pions,  up to $\mathcal{O}(g/f_{\pi}^3)$, as follows,
\bea
{\cal L}_1 &=&  -\frac{ 4g}{f_{\pi}^3} \epsilon^{\mu\nu\rho\sigma} ~{\rm Tr} [ 
V_{\mu} \partial_{\nu} \pi \partial_{\rho} \pi \partial_{\sigma} \pi]=-{\cal L}_2,  \label{WZW1} \\
{\cal L}_3&=&-\frac{2ig}{f_\pi} \epsilon^{\mu\nu\rho\sigma}\, {\rm Tr}[(\partial_\mu V_\nu)(V_\rho \partial_\sigma\pi-\partial_\rho\pi \,V_\sigma)] \nonumber \\ 
&&-\frac{4g^2}{f_\pi}\epsilon^{\mu\nu\rho\sigma}\, {\rm Tr}[V_\mu V_\nu V_\rho \partial_\sigma \pi].   \label{WZW2}
\eea
Then, the above new vector meson terms induce additional $3\rightarrow 2$ processes between the dark mesons.
and contribute to the $2\rightarrow 2$ self-scattering for dark mesons. In particular, the  $3\rightarrow 2$ annihilation cross section can be enhanced near the resonance \cite{resonance,average,simpmeson3} for $m_V\sim 3m_\pi$ or $m_V\sim 2m_\pi$, so the unitarity violation in the dark chiral perturbation can be delayed until higher energies.

We take the effective $3\rightarrow 2$ cross section before thermal average  \cite{average,simpmeson3} to be 
\be
(\sigma v^2)= \frac{\kappa b_V \gamma_V}{(\epsilon_V-u^2)^2+\gamma^2_V}
\ee
 where $b_V =  \frac{1}{4} (v_1^2+v_2^2 +v_3^2)^2-\frac{1}{2}(v_1^4+v_2^4+v_3^4)$, $\kappa$ is the velocity-independent coefficient depending on $m_\pi/f_\pi$ as well as $a$ and the anomaly coefficients, $c_1, c_2, c_3$,
and
\bea
(\epsilon_V,\gamma_V) =\left\{ \begin{array}{c}(\frac{m_V^2 - 4 m_{\pi}^2}{4 m_{\pi}^2},\frac{m_V \Gamma_V}{4 m^2_\pi}), \,\,\, \quad\quad m_V\approx 3m_\pi,   \\(\frac{m_V^2 - 9 m_{\pi}^2}{9m_{\pi}^2},\frac{m_V\Gamma_V}{9m^2_\pi}), \,\,\, \,\,\, m_V\approx 2m_\pi, \end{array}\right.
\eea
 and $u^2=\frac{1}{2}(v^2_1+v^2_2)-\frac{1}{4} v^2_3$  for two-pion resonances and $u^2=\frac{1}{3}(v^2_1+v^2_2+v^2_3)$ for three-pion resonances.  Here, $v_{1,2,3}$ are the speeds of initial dark pions given in the center of mass frame for the $3\rightarrow 2$ processes. 
Then, choosing the vector meson masses near the resonances and making the thermal average under the narrow width approximation with $\Gamma_V/m_V \ll 1$ where $\Gamma_V$ is the width of vector mesons,
we obtain the thermal averaged $3\rightarrow 2$ annihilation cross section \cite{average,simpmeson3} as
\begin{equation}
\langle \sigma v^2 \rangle_R \approx \left\{ \begin{array}{c} \frac{81 \pi}{128}\,\kappa \epsilon^4_V x^3 e^{-\frac{3}{2} \epsilon_V x},\,\,\, \quad\quad m_V\approx 3m_\pi, \\ \frac{8}{3}\sqrt{\pi} \, \kappa \epsilon^{3/2}_V x^{1/2}\, e^{-\epsilon_V x},\,\,\, \,\,\, m_V\approx 2m_\pi,  \end{array} \right..
\end{equation}
 As a result, the $3\to 2$ annihilation cross section can be enhanced at resonances of vector mesons, resolving the perturbativity problem of SIMP scenarios with dark mesons only. On the other hand, the $2\to 2$ self-scattering cross section is not enhanced at $m_V\sim 2m_\pi$ because of the overall velocity suppression for the corresponding resonance channel. Thus, we can keep the self-scattering cross section below the Bullet cluster bound, satisfying the correct relic density within the region of perturbativity.

\section{Conclusions}
We have presented the overview on dark mesons in a hidden QCD with dark flavor symmetries and focused on the interesting new candidates for dark matter with large self-interactions. 
The WZW 5-point interactions for dark mesons violate the $Z_2$ parity, so they lead to meson-number changing $3\to 2$ processes for dark mesons as in QCD. However, the dark mesons are indistinguishable and stable, thanks to the unbroken dark flavor symmetry, so they are good candidates for self-interacting dark matter.

We also sketched the roles of the dark local $U(1)'$ for maintaining the kinetic equilibrium for dark matter and avoiding the problem of structure formation at late times that would exist for a completely decoupled dark matter.  The dark flavor symmetry is partially broken due to the dark $U(1)'$, but the $3\to 2$ processes remain the dominant processes for determining the relic density as far as the mass splitting between mesons are small enough.

Finally, it was stressed that the excited states in the dark ChPT such as vector mesons can make the $3\to 2$ annihilation cross section enhanced near the new resonances and extend the parameter space for satisfying the unitarity beyond the dark ChPT.

\section*{Acknowledgments}

The work is supported in part by Basic Science Research Program
through the National Research Foundation of Korea (NRF) funded by the
Ministry of Education, Science and Technology (NRF-2019R1A2C2003738 and NRF-2018R1A4A1025334).

\end{document}